%% file: LRA_Lang_Zeng.tex
\documentclass[reprint,aip,jap,a4,showpacs,superscriptaddress,showkeys]{revtex4-1}%
\usepackage{amssymb}
\usepackage{amsfonts}
\usepackage{amsmath}
\usepackage{graphicx}%
\usepackage{color}
\usepackage[english]{babel}
\providecommand{\U}[1]{\protect\rule{.1in}{.1in}}

\begin{document}
\title[LRA Method for Efficient NEGF Calculation]{Low Rank Approximation Method for Efficient Green's Function Calculation of Dissipative Quantum Transport}
\author{Lang Zeng}
\email{langzeng@ime.pku.edu.cn}
\affiliation{Key Laboratory of Microelectronic Devices and Circuits, Institute
of Microelectronics, Peking University, 100871, P. R. China}
\affiliation{School of Electrical and Computer Engineering, Purdue University, West Lafayette,
USA 47907}
\affiliation{Network for Computational Nanotechnology, Purdue University, West Lafayette,
USA 47907}
\author{Yu He}
\affiliation{School of Electrical and Computer Engineering, Purdue University, West Lafayette,
USA 47907}
\affiliation{Network for Computational Nanotechnology, Purdue University, West Lafayette,
USA 47907}
\author{Michael Povolotskyi}
\affiliation{Network for Computational Nanotechnology, Purdue University, West Lafayette,
USA 47907}
\author{XiaoYan Liu}
\affiliation{Key Laboratory of Microelectronic Devices and Circuits, Institute
of Microelectronics, Peking University, 100871, P. R. China}
\author{Gerhard Klimeck}
\affiliation{School of Electrical and Computer Engineering, Purdue University, West Lafayette,
USA 47907}
\affiliation{Network for Computational Nanotechnology, Purdue University, West Lafayette,
USA 47907}
\author{Tillmann Kubis}
\email{tkubis@purdue.edu}
\affiliation{Network for Computational Nanotechnology, Purdue University, West Lafayette,
USA 47907}



\begin{abstract}
In this work, the low rank approximation concept is extended to the
non-equilibrium\ Green's function (NEGF) method to achieve a very efficient
approximated algorithm for coherent and incoherent electron transport. This new method is applied to inelastic
transport in various semiconductor nanodevices. Detailed benchmarks with exact NEGF solutions show 1) a very good agreement between approximated
and exact NEGF\ results, 2) a significant reduction of the required
memory,\ and 3) a large reduction of the computational time (a factor of speed up as high as $150$ times is observed). A non-recursive
solution of the inelastic NEGF transport equations of a $1000~\mathrm{nm}$
long resistor on standard hardware illustrates nicely the capability of this new method.

\end{abstract}

\date{\today}

\startpage{1}
\endpage{2}
\maketitle


\section{Introduction}%

\input{sec_introduction3.0.tex}


\section{Method\label{sec_II}}%

\input{sec_method3.1.tex}


\section{Results and Discussions\label{sec_III}}%

\input{sec_result5.1.tex}

\section{Conclusion\label{sec_IV}}%

\input{sec_conclusion2.1.tex}


\begin{acknowledgments}
Lang Zeng would like to thank China Scholarship Council (No.2009601231) for
financial support of his visiting study at Purdue University. The authors
would like to thank Peter Greck and Peter Vogl at the Walter Schottky
Institute and Akil Narayan at the Department of Mathematics at Purdue University for fruitful discussions.
Computational resources from nanoHUB.org
and support by National Science Foundation (NSF) (Grant No. EEC-0228390 and
No. OCI-0749140) are gratefully acknowledged. This work was also supported by the
Semiconductor Research Corporation's (SRC) Nano-electronics Research
Initiative and National Institute of Standards \& Technology through the
Midwest Institute for Nano-electronics Discovery (MIND), SRC Task 2141 and SRC Task 2273.
\end{acknowledgments}

\bibliographystyle{apsrev}
\bibliography{reference}

\end{document}

%% file: sec_introduction3.0.tex

Modern semiconductor devices have reached such small dimensions that carrier
confinement, interference effects and tunneling play an equally important role
as incoherent scattering, momentum and energy relaxation
do.~\cite{Moore_law,finfet_5nm,finfet_10nm,finfet_pmos} The non-equilibrium
Green's function (NEGF) method is among the most widely employed methods to
describe carrier dynamics in open quantum
systems.~\cite{kadanoff_book,Keldysh_1965} In fact, the NEGF method is applied
to a constantly growing variety of systems ranging from phonon
transport,~\cite{Xu_phonon,Mingo_phonon} spin
transport,~\cite{Green_T,Sergueev_spin}
electron dynamics in
metals,~\cite{Chen_andreev,Ke_metal,Schoen_quasi} organic
molecules\cite{Frederiksen_LOE} and fullerenes,~\cite{Sato_organic,Damle_2002,Li_C60,Schull_C60} and
semiconductor
nano-structures.~\cite{Register_QCL,Havu_2004,Lazzeri_hot_phonon_CNT,Luisier_transistor,Do_2006}
Unfortunately, the basic NEGF equations are numerically cumbersome and
extremely time demanding to solve. Therefore, several different
approximations for particular devices and situations have been developed to
reduce the numerical costs. The recursive Green's function method reduces the
peak numerical burden to a device dependent sub-block matrix of the
system's hamiltonian, and
the computational cost scales linearly with the number of blocks but cubically with the block size.~\cite{laker97,ferry1997}
It is widely used for the
simulations with one transport direction such as
FinFETs~\cite{finfet_5nm,finfet_10nm,finfet_pmos} and nanowire
structure.~\cite{wangjingJAP} Mode space approaches in similar wire structures as well as the newly developed Equivalent transport mode method separate the transport direction from transverse confinement
directions thus reducing the block size in each layer.~\cite{ModeSpacePRB,wangjingJAP,EM_method} All these methods usually require a
clear distinction between the transport direction and transverse degrees of
freedom. When this distinction gets blurred, as in the case of incoherent scattering, their numerical efficiency drops significantly. A very efficient method to solve ballistic NEGF\ equations is the
contact block reduction method (CBR).~\cite{CBR_JAP,CBR_JCE,CBR_PRB} However, this
method does not offer self-consistent incoherent scattering capability. Niche applications of the NEGF\ method have used sophisticated Wannier and Wannier
Stark functions to represent the transport problem in the presence of many
incoherent scattering mechanisms.~\cite{QCL_PRB,QCL_PSS} This specific basis
representation, however, is custom made for quantum cascade lasers and superlattices.

In this work, the low rank approximation (LRA) method is adapted to the
NEGF\ equations of electrons in simiconductors in the presence
of inelastic scattering on phonons.~\cite{LRA} This method is an extension of
the "basis reduction method" of Greck et al.~\cite{Greck_thesis,Greck_private} The concept of low rank approximation
is inherited from data modeling in control theory~\cite{control_theory}, machine learning~\cite{LRA}, signal processing~\cite{signal_processing},
bioinformatics~\cite{bioinformatic} for microarray data analysis etc. In the
framework of NEGF, the transport problem is transformed from the original basis
representation, i.e. in this work real space in effective mass hamiltonian presentation, to a more
appropriate basis of quasi-particle states that are close to the
quasi-particles of the actual device. In this representation, the number of
required basis functions is much less than in the original space, which allows
to reduce the numerical costs significantly. In so far, this method is closely
related to the beforementioned mode space approach. However, the LRA\ method
is a generalization to that, since it does not have any prerequisites to the
device geometry. In addition, the LRA\ implementation of this work uses a
third basis representation to enable real space defined inelastic scattering mechanisms.

In Sec.~\ref{sec_II}, the method of this work is introduced, its numerical complexity is
analyzed and differences of this method with existing approximations are
discussed. In Sec.~\ref{sec_III}, transport in homogeneous resistor and a resonant
tunneling diode is calculated. The comparisons of exact NEGF\ calculations with
the LRA\ approximated results show the accuracy of the presented method.
Limitations that this method (as every approximation approach does) faces are also
discussed in this section. To exemplify the computational strength of the
LRA\ method, electronic transport in a $1000~\mathrm{nm}$ resistor is
calculated in the end of this section. Energy resolved density spectrum illustrates
the transition from ballistic to drift diffusion transport in this resistor.
The paper concludes with a summary in Sec.~\ref{sec_IV}.

%% file: sec_method3.1.tex

\subsection{Low Rank Approximation Method\label{sec_IIA}}

NEGF calculations are time consuming since they involve the inversion and
multiplication of matrices with the rank $N$ of the system's hamiltonian. The
fundamental concept of the LRA method is to reduce the computational cost by
transforming the NEGF equations into a space of lower rank $n$ and solving
the equations therein. It is expected that the closer the basis functions of
the lower rank space are to the physically relevant quasi-particles of the
device, the better the LRA\ approximation is and the smaller the ratio $n/N$
can be chosen. The solution of the NEGF equations and all observables can be
transformed back into the original space after self-consistent calculation is achieved. In this way, the matrices that
represent the Green's functions and self-energies still have the lower rank
$n$, but the dimensionality $N$ which is required to maintain compatibility
with other equations (such as the Poisson equation) that might still be given
in the original space.

This method is exemplified on the stationary
vertical transport in laterally homogeneous quantum well hetero-structures that
are in contact with two charge reservoirs. The electron structure is
represented in terms of a single band effective mass hamiltonian $H_{0}$ that
is represented in a basis of $N$ position eigenfunctions%
\begin{equation}
H_{0}=\frac{-\hbar^{2}}{2}\frac{\mathrm{d}}{\mathrm{d}z}\frac{1}{m^{\ast
}\left(  z\right)  }\frac{\mathrm{d}}{\mathrm{d}z}+\frac{\hbar^{2}%
k_{\parallel}^{2}}{2m^{\ast}\left(  z\right)  }+V\left(  z\right)  ,
\end{equation}
where $k_{\parallel}$ is the in-plane electron momentum and $V\left(
z\right)  $ represents a position dependent potential. In the NEGF formalism,
stationary transport is determined by\ four coupled partial differential
equations
\begin{align}
G^{R} &  =\left(  E-H_{0}-\Sigma^{R}\right)  ^{-1},\nonumber\\
G^{<} &  =G^{R}\Sigma^{<}G^{R\dag},\nonumber\\
\Sigma^{<} &  =G^{<}D^{<},\nonumber\\
\Sigma^{R} &  =G^{R}D^{R}+G^{R}D^{<}+G^{<}D^{R}.\label{set1}%
\end{align}
Here, the electronic retarded and lesser Green's functions are given by
$G^{R}$, $G^{<}$, respectively.~\cite{tillmann_NEGF_intro,datta2005,datta1997}
$D$ is the sum of all environmental Green's functions that incorporate e.g.
phonons, and $\Sigma$ denotes the self-energies. The devices are in contact
with two charge reservoirs, represented with contact self-energies.~\cite{datta2005,datta1997}
If not explicitly stated otherwise, all calculations in this work include
inelastic scattering by longitudinal acoustic phonons given by the scattering
self-energies\cite{tillmann_NEGF_intro,tillmann_phonon_pe,kubis12}%
\begin{align}
& \Sigma_{\text{ac}}^{<,R}\left(  z,z^{\prime},k_{\parallel},E\right)
=\frac{1}{\left(  2\pi\right)  ^{3}}\frac{k_{B}TD^{2}_{ac}}{2\rho v_{s}^{2}%
}\nonumber\\
& \times\int\mathrm{d}\vec{q}_{\parallel}\mathrm{d}q_{z}\mathrm{e}%
^{iq_{z}\left\vert z-z^{\prime}\right\vert }\left[  \tilde{G}^{<,R}\left(
z,z^{\prime},\left\vert \vec{k}_{\parallel}-\vec{q}_{\parallel}\right\vert
,E+\hbar\omega_{q}\right)  \right.  \nonumber\\
& \left.  +\tilde{G}^{<,R}\left(  z,z^{\prime},\left\vert \vec{k}_{\parallel
}-\vec{q}_{\parallel}\right\vert ,E-\hbar\omega_{q}\right)  \right]
,\label{phonon_scattering}%
\end{align}
with the energy-averaged Green's functions%
\begin{equation}
\tilde{G}\left(  z,z^{\prime},q_{\parallel},E\right)  =\frac{1}{2\hbar
\omega_{D_{ac}}}\int_{E-\hbar\omega_{D_{ac}}}^{E+\hbar\omega_{D_{ac}}}\mathrm{d}E^{\prime
}G\left(  z,z^{\prime},q_{\parallel},E^{\prime}\right)  .
\end{equation}
The acoustic deformation potential and the material density is denoted by $D_{ac}$
and $\rho$, respectively. The acoustic phonon frequency is $\omega_{q}$ and
$v_{s}$ is the sound velocity.~\cite{tillmann_NEGF_intro,tillmann_phonon_pe,kubis12} The Debye frequency $\omega_{D_{ac}}$ limits the width of the average.

As the first step of this method, the $n$ eigenfunctions of the free particle
Hamiltonian $H_{0}$ with Neumann boundary conditions are solved that have
smallest eigen energy $E_{i}$%
\begin{equation}
H_{0}\phi_{i}=E_{i}\phi_{i},\text{ }i=1,2,\ldots n. \label{Schroedinger}%
\end{equation}
Hereby, $n$ is chosen such that the energy $E_{n}$ is about several $k_{B}T$ above
the highest chemical potential of all leads (with the Boltzmann constant
$k_{B}$ and the temperature $T$). Thereby, all quasi-particles with energies below $E_{n}$ are appropriately considered in the calculation.
This is essential to capture all occupied electronic states and to predict the density accurately. It is worth to mention that if only the transmission
around a given energy $E_{0}$ is required, it is sufficient to consider eigenstates of a few $k_{B}T$ around $E_{0}$.

In the second step, the $n$ orthonormal
eigenstates $\phi_{i}$ are set into the $n$ columns of a $N\times n$
dimensional matrix $S$. This matrix $S$ is unitary in the $n$ dimensional
space $\Omega$ spanned by the wavefunctions $\phi_{i}$, but note that it is not
unitary in the $N$ dimensional real space of step one%
\begin{equation}
S^{\dagger}S=I, \label{unity}%
\end{equation}%
\begin{equation}
SS^{\dagger}\neq I. \label{non_unity}%
\end{equation}
To define the locality/non-locality of scattering self-energies, the position
operator of the real space discretization $X$ is transformed into the reduced
rank basis in the third step%
\begin{equation}
X_{S}=S^{\dag}XS.
\end{equation}
The position operator $X$ is a diagonal matrix,
whereas the reduced rank matrix $X_{S}$ is a dense matrix. Therefore, the
operator $X_{S}$ is diagonalized to find the reduced rank position
eigenfunction basis $\left\{  \psi_{i}\right\}  $%
\begin{equation}
X_{S}\psi_{i}=x_{i}\psi_{i},\text{ }i=1,2,\ldots n. \label{position_eigen}%
\end{equation}
These orthonormal basis functions $\left\{  \psi_{i}\right\}  $ define the
columns of a squared, unitary $n\times n$ transformation matrix $P$. In the
basis $\left\{  \psi_{i}\right\}  $ the NEGF equations Eqs.~(\ref{set1}) read%
\begin{align}
G_{P}^{R}  &  =\left(  P^{\dag}T^{\dag}\left(  E-H_{0}\right)  TP-\Sigma
_{P}^{R}\right)  ^{-1},\nonumber\\
G_{P}^{<}  &  =G_{P}^{R}\Sigma_{P}^{<}G_{P}^{R\dag},\nonumber\\
\Sigma_{P}^{<}  &  =G_{P}^{<}D_{P}^{<},\nonumber\\
\Sigma_{P}^{R}  &  =G_{P}^{R}D_{P}^{R}+G_{P}^{R}D_{P}^{<}+G_{P}^{<}D_{P}^{R}.
\label{reduced_NEGF}%
\end{align}
Since every basis function $\psi_{i}$ is associated with a position $x_{i}$ the
equations above are discretized in a reduced rank real space representation. Position
dependent scattering self-energies such as the acoustic phonon scattering self-energy
in Eq.~(\ref{phonon_scattering}) are then self-consistently solved with the non-equilibrium
Green's functions in the numerically efficient reduced rank real space. The introduction of the reduced real space helps to avoid the back-transformation of the Green's functions into the original real space in the self-consistent calculation when position dependent scattering self-energy is calculated.

Once the NEGF equations Eqs.~(\ref{reduced_NEGF}) are converged, the diagonal and
the first off-diagonal elements of $G^{<}$ are transformed back into the original rank
$N$ real space representation. Observables such as the density or the current
density can then be evaluated in the original, high resolution real space. However, the
rank of the Green's functions in the $N$ dimensional system equals the
dimension of the space they are solved in, i.e. the rank equals $n$. The
smaller $n$ is compared to $N$, the more unreliable the $N$ dimensional
spatial information is, i.e. the stronger deviations of the LRA\ results from
the exact results are. It will be shown as one of the example results in the
next section, that the LRA approximated current density oscillates in the
original space, although the physical current of exact
calculations is conserved. To predict current voltage characteristics in the
LRA\ method, this inhomogenous current density is averaged over the device
excluding areas within $\Delta N/n$ from the leads (where $\Delta$ is the
average mesh point distance in the original real space representation).

\subsection{Comparison with existing efficient NEGF algorithms\label{comparison}}

It is important to highlight some differences of this method with other, well established efficient NEGF\ algorithms such as the CBR method~\cite{CBR_JAP,CBR_JCE,CBR_PRB}, mode space approaches~\cite{ModeSpacePRB,wangjingJAP} and recursive Green's function method.~\cite{laker97,ferry1997}

In the CBR method, the NEGF equations are first transformed into an
efficient representation to utilize the fact that ballistic calculations
require only some sections of the retarded Green's function $G^{R}$ to be
solved. A rectangular transformation that reduces the rank of the
NEGF\ equations is applied only after that first transformation. Although the
CBR method is very efficient, it is fundamentally limited to ballistic calculations.

The mode space approach assumes a separation Ansatz for the wave functions of
propagating quasi-particles. Typically, the Ansatz requires confined modes or
plane waves perpendicular to the transport direction. The mode space approach
allows a significant rank reduction of the NEGF equations. The computational burden is even further reduced if these modes are
well separated in energy and the particle propagation does not couple
different modes. However, if the device contains inhomogeneities
(impurities, non conformal confinement, etc.)\ the number of the modes is no longer a
conserved quantum number. Then, the modes are coupled and the rank of the mode space has to be large to predict
transport without loss of accuracy.

The recursive Green's function method allows to limit the calculation of the retarded Green's function to selected parameter intervals of the propagation space (i.e. sub-matrices of $G^R$, when $G^R$ is represented in matrix form). This allows limiting ballistic NEGF calculations on the required elements of $G^R$ only, which results in much faster transport solutions than the case when the complete $G^R$ is solved.~\cite{laker97} NEGF that includes incoherent scattering, however, requires the full $G^R$ which deteriorates the advantages of this recursive method.

In contrast to these three methods, the LRA\ method allows the inclusion of any
incoherent scattering as well as arbitrary device geometries. The "modes/sub-block matrix" of
the LRA\ method are device dependent wave functions that automatically include
non-conformal confinement - if such confinement appears.

\subsection{Numerical Complexity and Memory Usage Analysis\label{sec_IIC}}

\input{sec_numerical3.1.tex}

%% file: sec_numerical3.1.tex

Coherent quantum transport calculations for realistically extended devices have been shown to efficiently consume the computational power of over 220,000 processing cores.~\cite{OMEN_1,OMEN_2} Incoherent NEGF based calculations require about 100x more computational power, are limited to generally unrealistical small structures, and can only scale to about 100,000 cores.~\cite{Mathieu_phonon,OMEN_phonon} Involving incoherent scattering in realistically extended devices requires dramatically large computational resources.

The numerical complexity and memory usage of self-consistent NEGF calculations
reduce when the LRA\ method is applied. To qualify that, this section
compares the number of floating point operations and the memory usage of a "conventional" NEGF
calculation with an approximated ballistic NEGF solution that employs the LRA method. In the
following, the transport problem is assumed to be originally discretized with
$N$ orthogonal basis functions. Within the LRA\ method, the rank of the
NEGF\ equations is reduced down to $n$. The energy and other conserved quantum
numbers of the NEGF equations are resolved with a mesh of $N_{E}$ points. To
get a conserved current density within the self-consistent Born approximation,
$N_{i}$ iterations of the Green's functions and self-energies are required.
The integral in Eq.~(\ref{phonon_scattering}) is solved with $N_{ph}$ energy points.

\begin{description}
\item[Exact NEGF] The exact solution of one retarded Green's function involves
the inversion of a $N$ dimensional matrix which requires $O\left(
N^{3}\right)  $ floating point operations. The solution of each lesser Green's function
involves two matrix-matrix products with a numerical load of $O\left(
N^{3}\right)  $ floating point operations. The solution of the local scattering self-energy of Eq.~(\ref{phonon_scattering})
is $O\left(N_{ph} \times N\right)$ for each energy point in each iteration. In total, solving the NEGF\ equations exactly requires
$N_{E}\times N_{i}\times \left(O\left(  N^{3}\right)  +O\left(N_{ph} \times N\right)  \right)$ floating point operations,
while the memory needed to store the matrix representation of Green's
functions and self-energies is $N_{E}\times O\left(  N^{2}\right)  $ floating point numbers.

\item[Approximate NEGF] The LRA\ method can be decomposed into three steps: 1)
the transformation of the NEGF\ equations into the reduced space, 2) the
solution of the NEGF equations within the reduced space and 3) the
back-transformation of some relevant results into the original space. Step 1)
requires first to get the eigen states that construct the transformation
matrix, i.e. $O\left(  N^{2}n\right)  $ floating point operations. The memory used to store
the transformation matrix is $O\left(  Nn\right)  $ floating point numbers. The transformation of the
device's hamiltonian into the reduced space requires then $O\left(
N^{2}n\right)  $ floating point operations. The contact self-energies have to be transformed
for every energy point in every iteration. Since the contact self-energy is
zero except for the mesh points adjacent to the leads in the original space, each transformation requires
only $N_{E}\times N_{i}\times O\left(  n^{2}\right)  $ floating point operations and
$N_{E}\times O\left(  n^{2}\right)  $ floating point numbers to be stored. Solving the NEGF
equations in the reduced space in step 2) requires $N_{E}\times N_{i}\times
O\left(  n^{3}\right)  $ floating point operations and memory usage of $N_{E}\times O\left(  n^{2}\right)  $ floating point numbers.
The calculation of the acoustic phonon self-energy costs $O\left(N_{ph} \times n\right)$ operations in the reduce real space for each energy point in each iteration.
To calculate the energy resolved densities and current densities \cite{tillmann_NEGF_intro,tillmann_phonon_pe,tillmann_phonon_prb_2} in the original space, the step 3) requires to back-transform the diagonal and the
two first off-diagonals of $G^{<}$ of the original space. This transformation
requires $N_{E}\times O\left(  Nn^{2}\right)  $ floating point operations and memory usage of
$N_{E}\times O\left(  N\right)  $ floating point numbers.
\end{description}

In a typical effective mass NEGF calculation, the simulation setup reads $N=100$, $n=10$, $N_{E}=1000$, $N_{i}=10$ and $N_{ph}=10$. According to the analysis above, the numerical complexity of standard NEGF calculation in the typical effective mass situation is $O\left(
10^{10}\right)  $ floating point operations, and the memory usage is $O\left(  10^{7}\right)  $ floating point numbers; the
numerical complexity of LRA approximated NEGF calculation is
$O\left(  10^{7}\right)  $ floating point operations, and the memory usage is $O\left(  10^{5}\right)
$ floating point numbers. This observation demonstrates clearly that LRA method can reduce both numerical cost and memory usage significantly.

The comparison of the amount of floating point operations and memory usage between
the exact and the approximated LRA\ approach illustrate that the LRA method
offers approximated solutions of the NEGF\ equations much faster and with a
much smaller memory load than the exact solutions. In fact, one can easily find
NEGF equations of state of the art devices that are only solvable when the
LRA\ method is applied. To illustrate this, section \ref{1000nm} shows LRA approximated
NEGF\ results of a $1000~\mathrm{nm}$ homogeneous resistor with inelastic acoustic
phonon scattering calculated on single CPU.

%% file: sec_result5.1.tex

All devices in this section are laterally homogeneous layers grown in the
$z$-direction. Stationary transport along the $z$ direction is calculated for
conduction band electrons in the effective mass approximation. In the original
real space discretization (i.e. before LRA\ transformations are applied), the
Green's functions and self-energies are functions of two propagation
coordinates $z$ and $z^{\prime}$, the absolute in-plane momentum
$k_{\parallel}$ and the electron energy $E$. All devices of a given length $L$
are considered to be in contact with two charge reservoirs at $z=0$ and $z=L$, respectively.

\subsection{Homogeneous structure\label{50nm}}

Conduction band electrons of a $50~\mathrm{nm}$ thick, homogeneous layer of
GaAs with an effective mass of $m^{\ast}=0.067~m_{0}$%
~\cite{texbook} are
considered in this section. The NEGF\ equations are discretized with a
$0.5~\mathrm{nm}$ mesh spacing. The Fermi energies in both leads are assumed
to agree with the respective conduction band edge. The temperature is set to
$300~\mathrm{K}$. The conduction band in the device is set to be constant in
the first and last $5~\mathrm{nm}$ of the device and to drop linearly by the
amount of the applied bias voltage in the central $40~\mathrm{nm}$ of the device.

\begin{figure}[ptb]%
\centering
\includegraphics[
width=0.5\textwidth
]%
{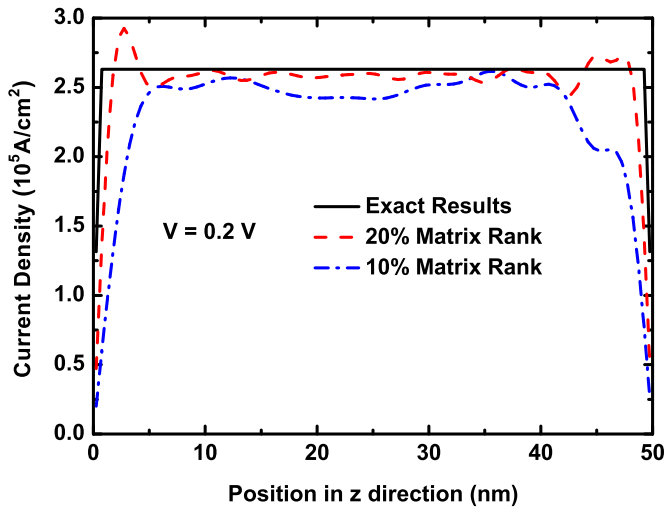}%
\caption{Spatially resolved current density in the homogeneous structure
described in the main text Sec.~\ref{50nm} with a linear potential drop of
$0.2~\mathrm{eV}$. The lines show result calculated with the NEGF\ method
solved exactly (solid) and with the NEGF\ method solved approximately with a
reduction of the matrix rank down to 20\% (dashed) and 10\% (dash-dotted).}%
\label{non_homo_current}%
\end{figure}

\begin{figure}[ptb]%
\centering
\includegraphics[
width=0.5\textwidth
]%
{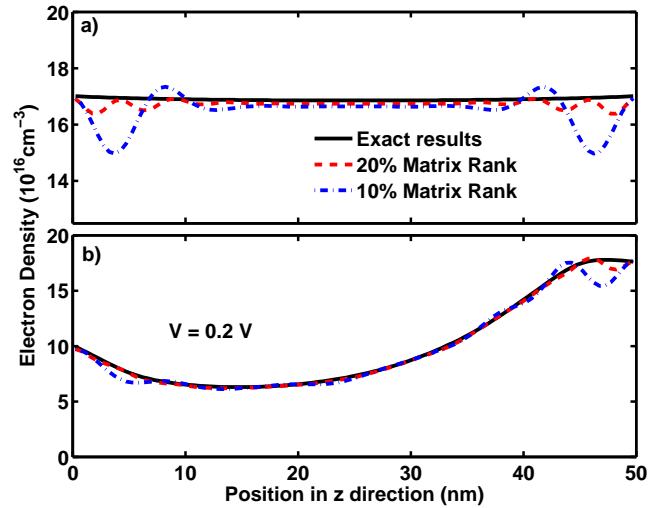}%
\caption{Calculated electron density of the homogeneous device of
Fig.~\ref{non_homo_current} in equilibrium (a) and when a linear potential
drop of $0.2~\mathrm{eV}$ is assumed (b). Results of an exact NEGF\ calculation
(solid) are compared with approximated NEGF\ calculations where the NEGF
equations' matrix rank is reduced down to 20\% (dased) and 10\%
(dash-dotted).}%
\label{homo_density}%
\end{figure}

Figure~\ref{non_homo_current} shows the spatially resolved current density
that results from an exact NEGF calculation as well as current densities of
LRA\ calculations when the matrix rank is reduced to $20\%$ and $10\%$ of the
original space. The exact calculation yields a
spatially constant current in the device, since inelastic phonon scattering is
included through a converged self-consistent Born
approximation. At the device boundaries the matrix elements of the exact contact self-energy is
non-zero as is common in the NEGF\ method. This non-vanishing self-energy allows
electrons to enter and leave the device, thus this contact self-energy
violates current conservation at the device boundaries. In the LRA\ method,
the contact self-energies are transformed into dense matrices. Their largest
elements are still located close to the device boundaries, which causes the
largest current fluctuations there. The larger the matrix rank reduction is,
the larger contact self-energy matrix elements within the device are.
Consequently, the larger the rank reduction is, the maximum amplitude of current
density fluctuations within the device is the higher. The smaller the rank of the
reduced real space is, the more dense the contact self-energies are. This allows electrons
to leave/enter the device at/to any device point in the reduced rank space. The non-constant
current densities in Fig.~\ref{non_homo_current} in the original real space indicate this
kind of violation of particle conservation.
Similar to the current fluctuations the density deviates from the exact
solution stronger, if the rank of the NEGF equations is reduced
more:\ Figure~\ref{homo_density} shows the electron density in the homogenous
layer of GaAs in equilibrium and at finite applied bias voltage. In both
cases, the deviations are strongest close to the leads.%

Both of the above figures indicate that the LRA\ method can reproduce exact
NEGF results in the device center up to close to the leads. This motivated the
device average of the current density described in Sec.~\ref{sec_IIA}. All
remaining current densities in this paper are such device averaged results.

\begin{figure}[ptb]%
\centering
\includegraphics[
trim=-0.027762in -0.236167in 0.027762in 0.236167in,
width=0.5\textwidth
]%
{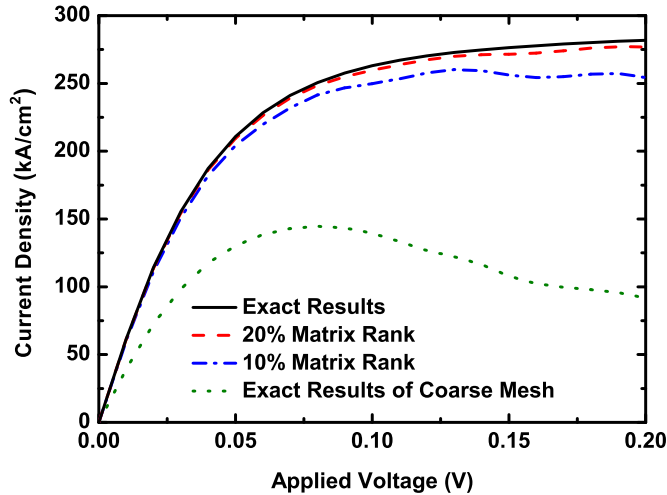}%
\caption{Comparison of I-V characteristic of the structure of
Fig.~\ref{non_homo_current} determined in exact NEGF\ calculations (solid) and
in approximated NEGF\ solutions where the NEGF\ matrix rank is
reduced to 20\% (dashed) and 10\% (dash-dotted). Also shown are the results
for an exact NEGF\ calculation of the same device when it is discretized with
a 10 times coarser real space mesh (dotted).}%
\label{homo_current}%
\end{figure}

Figure~\ref{homo_current} shows I-V characteristics of the $50~\mathrm{nm}$
thick homogeneous GaAs layer that have been calculated in the exact
NEGF\ method, as well as in the approximate LRA\ method with various rank
reduction levels. In addition, Fig.~\ref{homo_current} also shows an exact
NEGF calculation of the same device with a ten times coarser grid mesh.
Similar to previous figures, the deviation of the LRA approximated from the
exact NEGF results is larger, the larger the rank reduction is. Nevertheless,
a reduction of rank to $10\%$ is still able to well reproduce the
I-V\ characteristics, since only a small fraction of electronic states
in the low energy range contributes to the current density. In
contrast, results of exact NEGF calculations with a ten times coarser grid
deviate significantly from the full rank result: Such a coarser real space
mesh yields a different effective electron
dispersion~\cite{datta2005,datta1997} that deviates from the parabolic
dispersion within the relevant energies.

It is worth to mention that the boundary conditions for the electronic wave
functions of Eq.~(\ref{Schroedinger}) are relevant for the efficiency of the
LRA method. In agreement to similar findings of Mamaluy et
al.,~\cite{CBR_JAP,CBR_JCE,CBR_PRB} basis functions with Dirichlet boundary
conditions turned out to be inferior to Neumann conditions.

The LRA method and standard NEGF calculations were implemented in Matlab with 8 cores parallelization. For this concrete homogeneous structure calculation, the measured computational time for matrix rank reductions down to $10\%$ and $20\%$ is reduced by factories of $35\mathrm{X}$ and $87\mathrm{X}$ respectively compared to the full solutions, as listed in Table~\ref{Table1}. In our non-optimized LRA Matlab implementation, most of the time is spent to
transform Green's functions between different basis representations. If further optimization on matrix transformation is performed (as discussed in Sec.\ref{sec_IIC}), the factor of speed up can be even larger.

\begin{table*}[tbp] \centering
\begin{tabular}
[c]{|c|c|c|c|}\hline
& $\mathbf{10\%}$\textbf{ Matrix Rank} & $\mathbf{20\%}$\textbf{ Matrix Rank}
& \textbf{Exact Solution}\\\hline
\textbf{$\mathbf{50~\textbf{nm}}$ resistor} & $42\mathrm{s}/87\mathrm{X}$ & $104\mathrm{s}/35\mathrm{X}$ & $3658\mathrm{s}/1\mathrm{X}$\\\hline
\textbf{RTD structure} & $287\mathrm{s}/150\mathrm{X}$ & $1107\mathrm{s}/39\mathrm{X}$ & $43058\mathrm{s}/1\mathrm{X}$\\\hline
\textbf{$\mathbf{1000~\textbf{nm}}$ resistor} & too aggressive reduction & $459\mathrm{hr}$ & memory
exceed\\\hline
\end{tabular}
\caption{The measured time consumptions for the three examples in Sec.~\ref{sec_III}
for matrix rank reductions down to $10\%$ and $20\%$ as well as exact NEGF solutions are listed. The LRA method and standard NEGF calculations were implemented in Matlab with 8 cores parallelization.}%
\label{Table1}%
\end{table*}%

\subsection{Resonant tunneling diode\label{RTD}}

\begin{figure}[ptb]%
\centering
\includegraphics[
width=0.5\textwidth
]%
{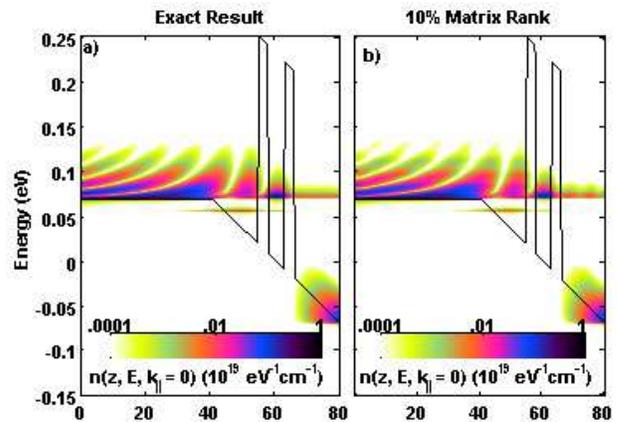}%
\caption{Conduction band profile (solid line) and contour plot of the energy
resolved electron density of the RTD structure described in the main text Sec.~\ref{RTD}. The
energy resolved density is calculated within the NEGF\ method exactly (a) and
approximately by a reduction of the NEGF\ equations' matrix rank to 10\% (b).
The filling of bound state in triangular well by inelastic acoustic phonon
scattering is well captured in the approximate method.}%
\label{rtd_density}%
\end{figure}

\begin{figure}[ptb]%
\centering
\includegraphics[
width=0.5\textwidth
]%
{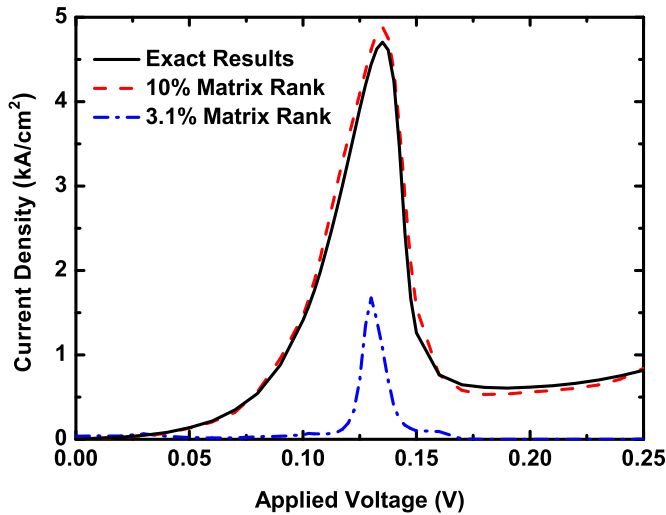}%
\caption{I-V characteristic of the RTD structure of Fig.~\ref{rtd_density}
calculated exactly (solid line) and approximately with 10\% (dashed) and 3.1\%
(dash-dotted) of the original NEGF\ equations' matrix rank.}%
\label{rtd_current}%
\end{figure}

This section explores the compatibility of the LRA\ method in quantum confined systems. The NEGF equations
are solved in a $80~\mathrm{nm}$ GaAs/Al$_{0.3}$Ga$_{0.7}$As resonant
tunneling diode (RTD) structure at $100~\mathrm{K}$. The RTD consists of two
$3~\mathrm{nm}$ wide Al$_{0.3}$Ga$_{0.7}$As barriers and a $5~\mathrm{nm}$
quantum well in the center. In addition, a $40~\mathrm{nm}$ flat band region
is located at emitter region. The effective mass for GaAs is $0.067~m_{0}$ and
$0.0919~m_{0}$ for Al$_{0.3}$Ga$_{0.7}$%
As.~\cite{tillmann_NEGF_intro,tillmann_phonon_pe,tillmann_self,texbook} The
band offset between these two materials is $230~\mathrm{meV}$%
.~\cite{tillmann_NEGF_intro,tillmann_phonon_pe,tillmann_self,texbook} In the
original real space representation, the device is discretized with a grid
spacing of $0.5~\mathrm{nm}$. The Fermi energies in the leads are set
$0.005~\mathrm{meV}$ beneath the respective conduction band edges. The
potential profile is assumed to be constant in the left most $40~\mathrm{nm}$
and to drop linearly in the remaining RTD region. This is illustrated by the
solid line in Fig.~\ref{rtd_density} (a) and (b) which show the same assumed
conduction band profile of the RTD in an exact NEGF calculation (a) and a
$10\%$ LRA approximated NEGF calculation (b). Figures~\ref{rtd_density}~(a)
and (b) also show contour graphs of the energy and spatially resolved electron
density of the RTD at vanishing in-plane momentum ($k_{\parallel}=0$)\cite{QCL_PRB,wacker_apl03,QCL_PSS,tillmann_NEGF_intro,tillmann_phonon_pe} when a
voltage of $0.1375~\mathrm{V}$ is
applied.
Both results agree very well: Fig.~\ref{rtd_density}~(b) deviates from (a)
only at the energy of about $0.07~\mathrm{eV}$ and positions $z\in\left[
65,80\right]  $. Even the confined state in the triangular quantum well locating at left
of the first RTD barrier (at energy of about $0.05~\mathrm{eV}$ and position
$50~\mathrm{nm}$) is well reproduced in the LRA calculation. This is
remarkable, since electrons can enter this state effectively only via
inelastic scattering. Therefore, inelastic scattering and tunneling are well
reproduced with the LRA method.
That can also be seen in Fig.~\ref{rtd_current}, which shows the I-V
characteristics of this RTD structure calculated in the exact NEGF method, as
well as in the LRA method with $10\%$ and $3.1\%$ of the original matrix rank.
Neither the current amplitude nor the resonance value get significantly
altered when the NEGF equations are solved with only $10\%\,$\ of the original
matrix rank. If the matrix rank is reduced too much, electronic states that
are relevant for the transport are neglected. Consequently, the current
density starts to deviate then. This is illustrated in Fig.~\ref{rtd_current}
with the I-V characteristic results of a LRA approximated NEGF calculation of only
$3.1\%$ of the original matrix rank. As stated in
Sec.~\ref{sec_IIA}, the ratio of the matrix rank reduction can be estimated
from the energy interval in which the states are occupied.%

For this RTD example, the measured computational time for matrix rank reductions down to $10\%$ and $20\%$ is reduced by factories of $39\mathrm{X}$ and $150\mathrm{X}$ respectively compared to the full solutions, as listed in Table~\ref{Table1}.

\subsection{1000nm long GaAs resistor\label{1000nm}}

\begin{figure}[ptb]%
\centering
\includegraphics[
width=0.5\textwidth
]%
{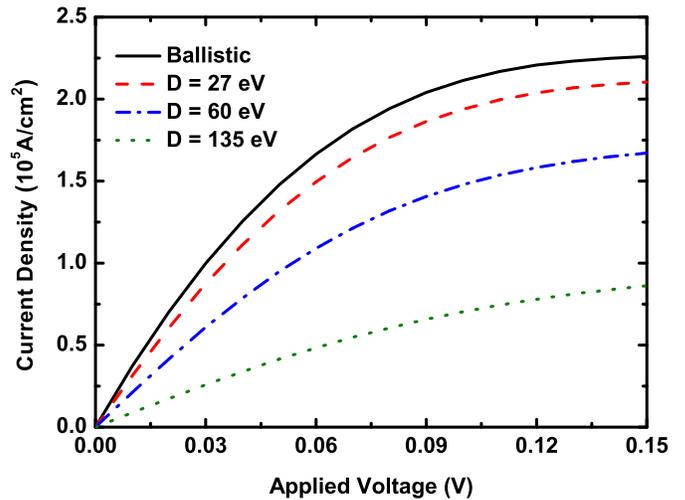}%
\caption{The I-V characteristic of the $1000$~$\mathrm{nm}$ homogenous
structure described in the main text Sec.~\ref{1000nm} when different values for the deformation
potential are used: $27~\mathrm{eV}$ (dashed), $60~\mathrm{eV}$ (dash-dotted) and
$135~\mathrm{eV}$ (dotted), and the ballistic results are shown as a solid curve. All results
are determined from approximated NEGF\ equations with a matrix rank of 20\% of
the original rank.}%
\label{long_current}%
\end{figure}

\begin{figure}[ptb]%
\centering
\includegraphics[
width=0.5\textwidth
]%
{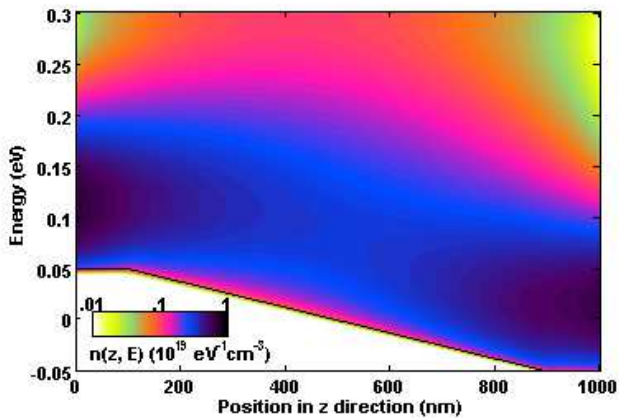}%
\caption{Energy resolved electron density of the 1000 nm long structure of
Fig.~\ref{long_current} with a deformation potential of
$135~\mathrm{eV}$ and a potential drop of $0.1~\mathrm{V}$.}%
\label{long_density}%
\end{figure}

The new LRA method is not only more efficient than the standard NEGF approach, it also opens up a space of device configurations that previously could not be tackled. This section considers
electronic transport in the presence of inelastic phonon scattering in a
$1000~\mathrm{nm}$ long homogeneous GaAs layer. In the range of
$100~\mathrm{nm}$ within the source and the drain contact/device interface, the
conduction band is assumed to be constant. In the remaining device, the
conduction band drops linearly according to the applied bias voltage. The
temperature of the phonon bath and the electrons in the leads is
$300~\mathrm{K}$ and the Fermi levels of the leads are set $0.1~\mathrm{eV}$
above the respective conduction band edge. The system is originally
discretized with a mesh spacing of $1~\mathrm{nm}$. The resulting NEGF
equations are approximated with a $20\%$ matrix rank. This reduces the numerical complexity of the NEGF\ equations such that
they have been solved on a single CPU\ without recursive approaches. The
nature of the transport is tuned from purely ballistic to almost drift
diffusion like by increasing the deformation potential $D$ of the phonon
scattering self-energy of Eq.~(\ref{phonon_scattering}). Hereby, three
different scattering potentials have been considered:\ $27~\mathrm{eV}$,
$60~\mathrm{eV}$ and $135~\mathrm{eV}$, which corresponds to a scattering rate
of $1\times10^{12}~\mathrm{s}^{-1}$, $5\times10^{12}~\mathrm{s}^{-1}$ and
$2.5\times10^{13}~\mathrm{s}^{-1}$ for electrons with kinetic energy of
$0.3~\mathrm{eV}$. The impact of the scattering is illustrated in
Fig.~\ref{long_current} as it shows the calculated I-V characteristics of the
device with various scattering strengths. The I-V characteristic is almost
ohmic in the case of a deformation potential of $135~\mathrm{eV}$.%
The nature of transport at this large deformation potential can be understood
from Fig.~\ref{long_density}. It shows the energy and spatially resolved
electron density for the $1000~\mathrm{nm}$ long resistor in the case of
$0.1~\mathrm{V}$ applied bias voltage. Electrons that originate from the
source contact propagate about $400~\mathrm{nm}$ in the device before they
start to significantly dissipate energy. Then, however, these electrons follow
the potential drop of the device and thereby start to maintain a local
equilibrium distribution. In this way, the electrons experience a transition
from effectively ballistic transport into the drift diffusion of the rightmost
$500~\mathrm{nm}$ of the device.

%% file: sec_conclusion2.1.tex

In this work, the low rank approximation method is applied to efficiently and
accurately solve the approximated NEGF equations in the effective mass approximation. It
is shown that this method reliably solves the electronic transport in the
ballistic and incoherently scattered transport regime. Quantum effects that
are natively included in the NEGF\ equations (such as interferences,
confinement and tunneling) are accurately reproduced by the LRA method, but
with a fraction of the numerical load of the original NEGF equations. This
method differs from existing NEGF approximations since it allows to include
incoherent scattering (in contrast to the CBR\ method) and does not require
specific device shapes (in contrast to the mode space approach and EM method).

In this paper, the LRA method has been applied to homogeneous resistors and
resonant tunneling diodes, i.e. to classical resistors and quantum confined
structures. In both systems, a very good agreement of the LRA-\ approximated I-V
characteristics and energy and spatially resolved densities with exact
NEGF\ solutions has been demonstrated. Significant deviations of the
LRA\ method from exact results appear only for matrix rank reductions that
are too strong and neglect states relevant for transport. To show the
efficiency and power of the LRA method, transport in a $1000~\mathrm{nm}$ long GaAs
resistor has been calculated. An exact NEGF\ calculation of this long device
is not feasible without recursive algorithms. The LRA\ method, however,
allowed to solve this system without recursion and even when incoherent
scattering was increased to almost drift diffusion like transport.